# Angle-resolved photoemission spectroscopy of superconducting $(La,Pr)_3Ni_2O_7/SrLaAlO_4$ heterostructures

Peng Li[1,2†], Guangdi Zhou[1,2†], Wei Lv[2†], Yueying Li[2†], Changming Yue[1,2,3*], Haoliang Huang[1,2], Lizhi Xu[2], Jianchang Shen[4,5], Yu Miao[4,5], Wenhua Song[2], Zihao Nie[2], Yaqi Chen[2], Heng Wang[2], Weiqiang Chen[1,2], Yaobo Huang[6], Zhen-Hua Chen[6], Tian Qian[7], Junhao Lin[1,2], Junfeng He[4,5*], Yu-Jie Sun[1,2*], Zhuoyu Chen[1,2*], Qi-Kun Xue[1,2,8]

[1]Quantum Science Center of Guangdong-Hong Kong-Macao Greater Bay Area, Shenzhen 518045, China

[2]State Key Laboratory of Quantum Functional Materials, Department of Physics, and Guangdong Basic Research Center of Excellence for Quantum Science, Southern University of Science and Technology, Shenzhen 518055, China

[3]Guangdong Provincial Key Laboratory of Advanced Thermoelectric Materials and Device Physics, Southern University of Science and Technology, Shenzhen, 518055, China

[4]Department of Physics and CAS Key Laboratory of Strongly-coupled Quantum Matter Physics, University of Science and Technology of China, Hefei 230026, China

[5]Hefei National Laboratory, University of Science and Technology of China, Hefei 230088, China

[6]Shanghai Synchrotron Radiation Facility, Shanghai Advanced Research Institute, Chinese Academy of Sciences, Shanghai 201204, China

[7]Beijing National Laboratory for Condensed Matter Physics and Institute of Physics, Chinese Academy of Sciences, Beijing 100190, China

[8]Department of Physics, Tsinghua University, Beijing 100084, China

[†]These authors contributed equally.

[*]E-mail: chenzhuoyu@sustech.edu.cn, sunyj@sustech.edu.cn, jfhe@ustc.edu.cn, yuecm@sustech.edu.cn

**Ruddlesden-Popper bilayer nickelate thin film superconductors discovered under ambient pressure enable vast possibilities for investigating electronic structures of the superconducting state. Here, we report angle-resolved photoemission spectroscopy (ARPES) measurements of 1, 2, and 3 unit-cell epitaxial $La_{2.85}Pr_{0.15}Ni_2O_7$ films grown on $SrLaAlO_4$ substates, through pure-oxygen *in situ* sample transportation. Evidence obtained using photons with distinct probing depths shows that conduction is localized primarily at the first unit cell near the interface. Scanning transmission electron microscopy (STEM), together with energy-dispersive X-ray spectroscopy (EDS) and electron energy loss spectroscopy (EELS), indicates that interfacial Sr diffusion and pronounced *p-d* hybridization gradient may collectively account for the interfacial confinement of conduction. Fermi surface maps reveal hole doping compared to non-superconducting ambient-pressure bulk crystals. Measurements of dispersive band structures suggest the contributions from both Ni $d_{x^2-y^2}$ and $d_{z^2}$ orbitals at the Fermi level. Density functional theory (DFT) + U calculations capture qualitative features of the ARPES results, consistent with a hole-doped scenario. These findings constrain theoretical models of the superconducting mechanism and suggest potential for enhancing superconductivity in nickelates under ambient pressure.**

## Introduction

High-temperature superconductivity has been a central topic of condensed matter physics for decades [1-3], particularly within the cuprate family, where the superconducting mechanism is predominantly associated with the $d_{x^2-y^2}$ single-band Fermi surface observed by angle-resolved photoemission spectroscopy (ARPES) [4-6]. Although Ni and Cu are distinct elements, infinite-layer nickelates, a family of superconductors discovered in 2019, exhibit a lattice structure analogous to that of cuprates and share a similar nominal $3d^9$ electron occupancy [7-9]. ARPES studies of infinite-layer nickelates have confirmed that, despite the presence of rare-earth $5d$ orbitals crossing the Fermi level, the electronic structure is fundamentally governed by $d_{x^2-y^2}$ bands, again similar to the cuprates [10,11]. In contrast, bilayer nickelates, discovered in 2023 to be superconducting under high pressure over 14 GPa with transition temperature near 80 K, while also comprising Ni, exhibit an average $3d$ electron occupancy of around 7.5, markedly different from the $3d^9$ configuration in cuprates [12-16]. Furthermore, bilayer nickelates feature apical oxygen atoms between the superconducting Ni-O layers, a structural characteristic absent in cuprates. This apical oxygen induces coupling between the two Ni-O layers via the $d_{z^2}$ orbitals, prompting theoretical debates regarding the precise role of these orbitals in the superconducting mechanism [17-34]. Critical aspects of the debates involves whether the $d_{z^2}$-dominant bands intersect the Fermi level and whether the superconducting state is hole-doped or electron-doped relative to the non-superconducting ambient-pressure parent. ARPES measurements on non-superconducting bulk bilayer nickelates at ambient pressure demonstrated that the $d_{z^2}$ bands lie below the Fermi level [35,36]. However, high-pressure condition that is necessary for superconductivity is incompatible with photoemission experimental techniques, and the presence of mixed phases further complicates the analysis [37-40], leaving the electronic structure of the superconducting state undetermined.

Recently, ambient-pressure superconducting bilayer nickelate ultrathin films were reported [41,42], offering an opportunity for investigating their electronic structures. Unlike bulk superconducting samples subjected to near-isotropic external pressure, the thin films experience biaxial epitaxial strain from the substrate, resulting in an in-plane lattice compression and out-of-plane lattice elongation. This distinct lattice deformation may lead to variations in electronic structures, differing from the assumptions of many existing theoretical and computational studies that are based on high-pressure, isotropic compression [43,44]. As such, the strained bilayer nickelate ultrathin films provide a unique platform for exploring the high-temperature superconducting mechanisms. In this work, we have performed ARPES measurements on these recently discovered bilayer nickelate superconducting ultrathin films, providing initial evidence to the open questions listed above.

## Results

Figure 1 shows the magnetotransport measurements and ARPES Fermi surface maps of one unit-cell (1UC), 2UC, and 3UC $La_{2.85}Pr_{0.15}Ni_2O_7$ films on treated (001)-oriented $SrLaAlO_4$ substrates, epitaxially grown and annealed with same conditions. Samples with all different thicknesses exhibit superconducting transition in transport measurements (Figs. 1d-1f), though with notable thickness-dependent variations. Degradation of superconducting properties in 2UC and 1UC samples is attributed to the inevitable oxygen loss during the relatively slow cooling process (300 K to 200 K, cooling rate limited by the cryogen-free cryostat) under vacuum. This oxygen loss is increasingly significant with reduced film thickness. With existence of interfacial scattering (e.g. step terraces from substrate miscut angles and interdiffusions), the reduced dimensionality leads to disproportionately severe degradation of macroscopic transport characteristics in the monolayer limit. Nonetheless, these transport measurements confirm that the ARPES measurement conducted at 10 K reflect the superconducting state across 1UC to 3UC samples. For ARPES measurements, samples were transported in a 0.2-MPa pure oxygen suitcase from the growth chamber to the synchrotron end station, and then were rapidly quenched to < 100 K on the manipulator (temperature constantly set to 10 K). This protocol minimizes oxygen loss compared to transport measurement procedures, particularly critical for the ultrathin 1UC sample.

Linear horizontally (LH) polarized photons of 200 eV in the experimental geometry as defined in Fig. 1j maximize the capability of probing photoelectrons from different electron orbitals (Figure S1). The Fermi surface maps exhibit similar topologies across the three heterostructures, while the 2UC and 3UC samples display more smeared features compared to that of the 1UC sample (Figs. 1g-1i, constant energy contours in Figure S2). By extracting the momentum distribution curves (MDCs) along MXM (cut 1 in Fig. 1g) for the three films (Fig. 1k), we observe that the peak separations are similar across the films, suggesting the absence of significant chemical potential shifts across samples with varying film thickness. Notably, these peak separations are smaller than that of the extracted MDC from bulk $La_3Ni_2O_7$ non-superconducting compound at ambient pressure along the same direction[3], indicating hole doping in the superconducting films (Fig. 1l). Weak intensity is observed at the Γ point, suggesting that the Fermi level may be nearly touching a band bottom at this location. Moreover, the diffuse yet pronounced intensity around the M point suggests the presence of an additional band. The diffuse appearance of these features may stem from the underlying nature of this band, which will be discussed later.

Before presenting detailed band structure measurements and analysis, we first investigate the origin of smearing for 2UC and 3UC samples' Fermi surface measurements, by characterizing the electronic structures across the film thickness. The integrated energy distribution curves (EDCs) obtained from 200 eV synchrotron photons (Fig. 2a) and 21.2 eV Helium-I photons (Fig. 2b) are compared. The photoelectron mean free path with 21.2 eV incident photons is significantly smaller than that with 200 eV photons, resulting in a much shorter probing depth and much stronger surface sensitivity. While all three films (1UC, 2UC, and 3UC) exhibit metallic behavior, as indicated by integrated EDCs crossing the Fermi level under 200 eV illumination, distinct features are observed for each film when measured with 21.2 eV photons: in contrast to the 1UC sample, which retains significant amount of electron density of states near the Fermi level, the 2UC sample shows a substantial suppression of intensity in this region, and the 3UC sample exhibits an absence of intensity near the Fermi level, suggesting an insulating-like behavior (intensities are normalized at higher binding energies, see Fig. 2b inset). The thin insulating layer allows any excess charge to tunnel away efficiently, thus no obvious charging effect is observed (Figure S3). A plausible scenario involves a pronounced conductivity gradient across the film thickness, as shown in Fig. 2c, where the 1st UC near the interface is highly conductive, while the 3rd UC is almost insulating.

The origin of such significant conductivity gradient may be elucidated through scanning transmission electron microscopy (STEM) techniques (Fig. 2d and insets, full-range image in Fig. S4). Atomically resolved energy-dispersive X-ray spectroscopy (EDS) reveals Sr diffusing across the interface with the diffusion length ~1UC. Since Sr has a +2 valence, its substitution for the +3 valence La (or Pr) induces hole doping. This implies that Sr concentrations, on top of the substrate epitaxial compressive strains, may help the emergence of superconductivity. Additionally, electron energy loss spectroscopy (EELS) analysis of the oxygen K edge pre-peak intensity, a signature of *p*-*d* hybridization [45,46] (Fig. 2e), exhibits marked gradient across the film thickness (Fig. 2f). One possible scenario is that oxygen loss at the surface induces insulating behavior at the top layer [41,42], and the resulting oxygen content variations throughout the film leads to the observed *p*-*d* hybridization gradient [34]. The existence of top layers may have protected the 1st UC from significant oxygen depletion. Both Sr diffusion, which may induce hole doping at the interface, and oxygen loss, which may cause reduced *p*-*d* hybridization at the surface, may collectively contribute to the confinement of superconductivity near the interface in the measured $La_{2.85}Pr_{0.15}Ni_2O_7/SrLaAlO_4$ system. In this context, improvement of experimental technique is needed to control and maintain oxygen content across the films during the sample growth, storage, and transfer, thereby enhancing superconducting properties, particularly in the top layers. Given the similar Fermi surface topologies among 1UC, 2UC, and 3UC samples and the higher spectroscopic quality of the 1UC, we focus mostly on the 1UC below for band structure analysis.

Figures 3a display the Fermi surface maps of the 1UC film in LV polarization (constant energy contours in Figure S5). Due to the matrix element effect (Figure S1) of ARPES, these maps clearly differ from those in LH polarization shown in Fig. 1. Figure 3b exhibits summed Fermi surface intensity obtained from both LH and LV polarized photons, which contains more comprehensive orbital information. We reproduce a schematic of the orbital-characterized Fermi surface map [6,47] in Fig. 3c, showing contributions from both $d_{x^2-y^2}$ and $d_{z^2}$ orbitals (denoted as α, β, and γ pockets). Based on area integrals of this Fermi surface map, we estimate a hole doping level of 21% ± 5% (i.e. 0.21 holes per Ni) relative to that of the non-superconducting parent bulk, assuming the two-dimensional nature of the electronic structure [35] (reproduced in Fig. 1i). The second derivative spectra exhibit band dispersions for both β and γ pockets crossing the Fermi level in the first and the second Brillouin zones (Figs. 3d & 3e). The quantitative determination of the band dispersions near the Fermi level (Fig. 3f) can be obtained from Figs. 3g-3l. The small peak in the Fig. 3j MDC at the Fermi level near the Γ point may be corresponding to the same weak intensity observed in Fermi surface map (Fig. 1d), implying the existence of a band bottom right above Fermi level. The band dispersions (Figs. 3g and 3h) and the corresponding MDCs (Figs. 3j and 3k) confirms the presence of the γ pocket around M point crossing the Fermi level.

Density functional theory (DFT) + U calculations of a nickelate bilayer based on strained thin film lattice structures [42] without Sr substitution are presented in Fig. 4. The larger energy window in Fig. 4a displays the distribution of the $d_{x^2-y^2}$ and $d_{z^2}$ orbital related bands. Fig. 4b is a zoom-in version of the same set of data, deliberately plotted with the similar energy window and high-symmetry path as in Fig. 3f. Note that the Fermi level is calibrated for comparison with the experiment, based on the proximity of a band minimum to the Fermi level at the Γ point. The Fermi level for the undoped case is also shown for reference. The calculated band dispersions qualitatively capture the ARPES results. The quantitative differences hint the need for strongly correlated algorithms [48]. The diffuse manifestation of the $d_{z^2}$ Fermi surface observed in ARPES may also arise from electron correlations in this $d_{z^2}$ band, which features a partially flat region above the Fermi level in DFT calculations. Figures 4c and 4d present the Fermi surfaces for undoped and hole-doped cases, with the latter qualitatively agrees with experiments (Fig. 3e) qualitatively. Fig. 4e is the associated orbital dependent density of states.

## Conclusion

In summary, angle-resolved photoemission spectroscopy measurements indicate that the superconducting state in $La_{2.85}Pr_{0.15}Ni_2O_7/SrLaAlO_4$ heterostructures is hole-doped, with contributions from both $d_{x^2-y^2}$ and $d_{z^2}$ orbitals at the Fermi level. Photoemission spectroscopy, together with atomically resolved energy-dispersive X-ray spectroscopy and electron energy loss spectroscopy, indicates a combined effect of interfacial Sr diffusions and surface suppressions of *p-d* hybridization, confining conduction near the interface. These findings provide an initial basis for further exploration of the superconducting mechanisms in bilayer nickelates and offer insights into strategies for enhancing their superconducting properties.

## Methods

**Samples.** High quality $La_{2.85}Pr_{0.15}Ni_2O_7$ thin films with different thickness were grown using the Gigantic-Oxidative Atomically Layer-by-Layer Epitaxy (GOALL-Epitaxy) method [49], followed by a 30-minute post annealing at 575°C within the growth chamber. The $SrLaAlO_4$ (001) substrates (MTI, China) were annealed by placing a $LaAlO_3$ substrate face to face with a $LaSrAlO_4$ substrate under atmospheric conditions at 1030 ℃ or 1080 ℃ for 2 hours. Platinum electrodes are deposited on the substates/film before ozone annealing for grounding. The grown samples were transferred to synchrotron beamline station and helium-lamp ARPES station in suitcases, filled with oxygen at ~0.2 MPa. The purity of oxygen is to 99.999% and the base-pressure of the suitcases were maintained below $1\times10^{-8}$ mbar prior to filling with oxygen.

**ARPES measurements.** Synchrotron-ARPES measurements were performed at beamline BL09U of Shanghai Synchrotron Radiation Facility (SSRF) in China and the overall energy resolution was set to be better than 20 meV at 200 eV photon energy and the angular resolution is ~0.2 degree for measurements. Base pressure of the beamline station is better than $4\times10^{-11}$ Torr. Helium-lamp (21.2 eV) ARPES measurements were performed at a lab-based ARPES system with a base pressure better than $5\times10^{-11}$ Torr. After testing various photon energies from different light sources, 200 eV was found to provide optimal resolution for band dispersions. During ARPES measurements, the remaining samples were kept under oxygen environment at ~0.2 MPa. Sample transfers in ultrahigh vacuum were conducted as quickly as possible to minimize surface oxygen loss. Measurement temperature is ~10 K.

**Band structure calculations.** A simple thin film crystal structure is constructed considering only half unit-cell of $La_3Ni_2O_7$ using the lattice constants, Ni-O-Ni angles and distances and La-La distance determined experimentally [42]. The actual length of vacuum in the calculation is more than 30 Å. The positions of oxygens were relaxed while fixing the lattice constants and positions of La and Ni atoms using the VASP package [50-52]. The band structures are calculated using the DFT+U method with preliminary parameters U=5 eV and J=1 eV. The tight-binding (TB) model Hamiltonian for the bands near the Fermi energy is obtained from the maximally-localized Wannier function method implemented in the package of wannier90 [53,54]. A 125×125 k-point mesh grid in the first

Brillouin zone is used to calculate the iso-energy surfaces for different binding energies based on the TB model Hamiltonian.

**Scanning transmission electron microcopy (STEM).** The specimens for the cross-section STEM were prepared utilizing the focused ion beam (FIB) technique on a FEI Helios G4 HX dual-beam FIB/SEM machine. The HAADF images were photographed using a double spherical-aberration corrected FEI Themis Z system at 200 kV, with a high-brightness field-emission gun (X-FEG) equipped with a monochromator installed onto this microscope. The probe semi-convergence angle is 25 mrad and the collection angles for the STEM images ($\beta1$ and $\beta2$) were 90 and 200 mrad, respectively. The EDS mappings were obtained on the Super X FEI System in STEM mode. The beam currents of 40 and 136 pA were used for HAADF imaging and EDS analyses, respectively. EELS spectra of the O-K edge were performed on a FEI Themis Z equipped with a high stability stage, a Gatan Quantum ER/965 spectrometer, operated at 200 kV with 200 pA beam current. The probe currents are controlled less than 200 pA to minimize radiation damage with the scan parameters used: ~1-Å probe, 9-$Å^2$ scan pixel size (applied 16×16 sub-pixel scan), and 300 μs per pixel dwell time. Before extraction and thinning, platinum and carbon layers were deposited by electron beam and ion beam respectively to prevent ion beam damage of the sample surface. All operations were done at room temperature.

**Transport measurements.** Electric transport measurements were performed in a closed-cycle helium-free cryostat with a base temperature of approximately 1.8 K and a magnetic field of up to 14 T. Platinum Hall bar electrodes were fabricated on 5 × 5 $mm^2$ samples using magnetron sputtering through a pre-patterned hard shadow mask. Prior to sample loading, the sample chamber was cooled to 275 K, followed by helium gas purging to remove residual air. The final pressure in the chamber was maintained at approximately 10 Torr. To minimize oxygen loss, the sample was cooled below 200 K in 10 minutes, to minimize the time under conditions that leads to oxygen loss.

## Acknowledgements

We acknowledge the discussions with Hongtao Yuan, assistance from Shuting Peng during the ARPES measurements, and the support from International Station of Quantum Materials.

## Funding

This work was supported by the National Key R&D Program of China (2024YFA1408101 and 2022YFA1403101), the Natural Science Foundation of China (92265112, 12374455, & 52388201), the Guangdong Provincial Quantum Science Strategic Initiative (GDZX2401004 & GDZX2201001), the Shenzhen Science and Technology Program (KQTD20240729102026004), and the Shenzhen Municipal Funding Co-Construction Program Project (SZZX2301004 & SZZX2401001). Yueying Li acknowledges the support by China Postdoctoral Science Foundation (2024M761276). Changming Yue acknowledges the support by National Natural Science Foundation of China (1247041908) and Guangdong Provincial Key Laboratory of Advanced Thermoelectric Materials and Device Physics (2024B1212010001). Jianchang Shen, Yu Miao, Junfeng He acknowledges support by the National Key R&D Program of China (2024YFA1408100), the International Partnership Program of the Chinese Academy of Sciences (123GJHZ2022035MI), and the Innovation Program for Quantum Science and Technology (2021ZD0302802). Tian Qian acknowledges the support by National Key R&D Program of China (2022YFA1403800) and Natural Science Foundation of China (U22A6005).

## Author contributions

Q.K.X. supervised the entire project. Z.C. initiated the study and coordinated all the research efforts. P.L. and Y.L. led synchrotron-ARPES measurements, with assistance from L.X., J.S., Y.M., and W.S., and under supervision from Z.C., J.H. and Y.J.S. G.Z. and W.L. prepared samples customized for ARPES measurements, with assistance from Z.N. and Y.C. C.Y. performed DFT+U calculations. H.H. and J.L. provided STEM, EDS, and EELS analysis. J.S. and Y.M. performed helium-lamp-ARPES measurements under supervision from J.H. H.W. conducted oxygen loss analysis. W.C. provided theoretical support. Y.H. and Z.H.C. supported ARPES measurements at the beamline. T.Q. provided beamline resources and participated discussions. Z.C., P.L., and Y.L. wrote the manuscript with input from C.Y., J.H., Y.J.S., and all other authors.

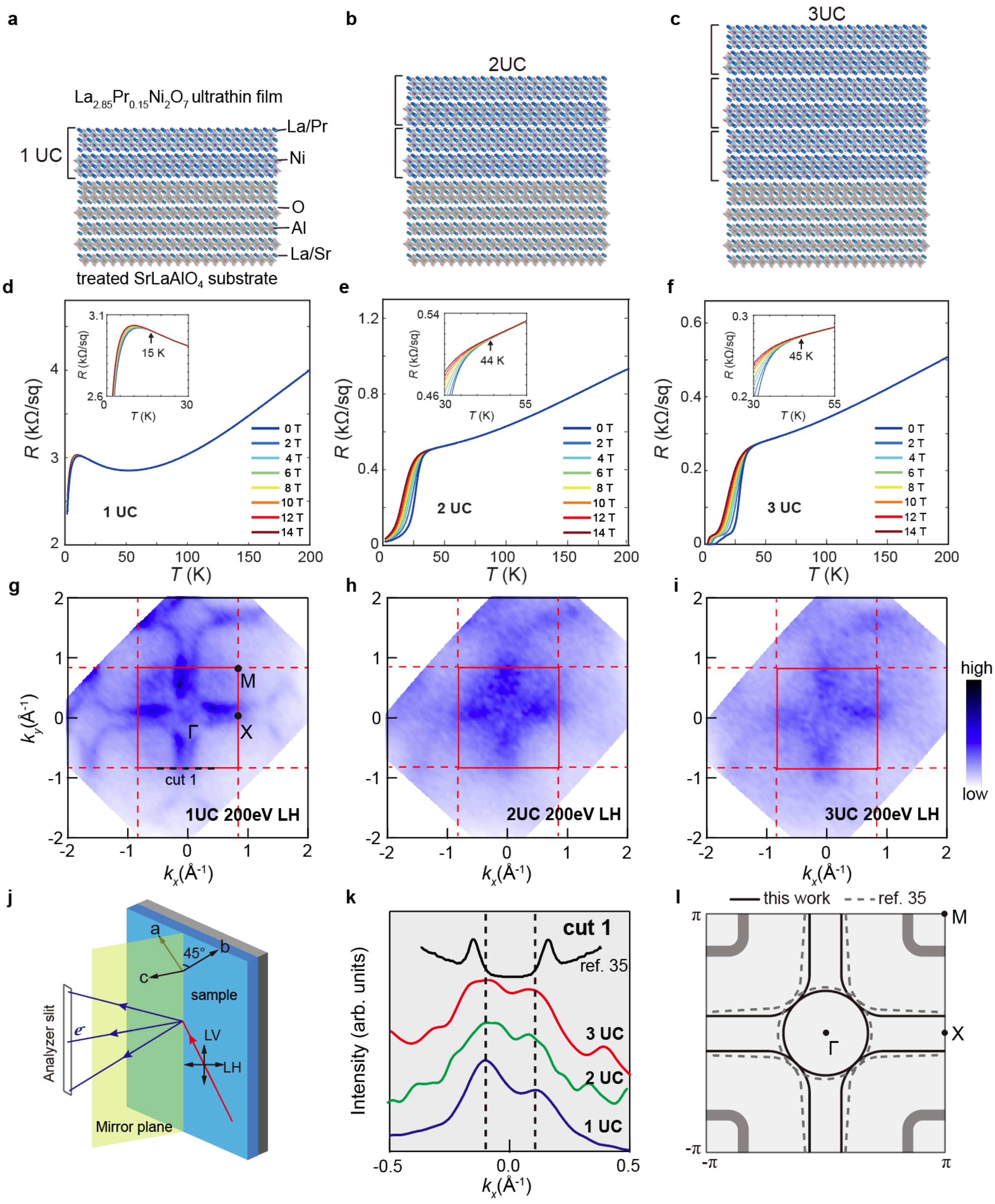

**Figure 1 | Fermi surface maps of $La_{2.85}Pr_{0.15}Ni_2O_7$/$SrLaAlO_4$ heterostructures. a-c,** Structural schematics of $La_{2.85}Pr_{0.15}Ni_2O_7$ films with thickness of 1 unit cell (1UC), 2UC and 3UC, respectively, grown on treated $SrLaAlO_4$ substrates. **d-f,** Magnetotransport measurements in 1UC, 2UC and 3UC films, with insets zoomed in the onset region. Arrows indicate the onset temperatures of magnetic field responses. Note that surface oxygen loss during cooling down in vacuum causes degradation of superconductivity in 1UC and

2UC samples. To reduce oxygen loss, the 1UC sample for transport was capped with around half nm $SrTiO_3$ after growth, while 2UC and 3UC samples are measured without capping. **g-i** Fermi surface maps of 1, 2 and 3UC thin films measured via angle-resolved photoemission spectroscopy (ARPES), using linear horizontally polarized (LH) photons with an energy of 200 eV. The red solid squares denote the first Brillouin zone (BZ) with high symmetry points Γ, M, and X marked in panel **g**. The regions enclosed by the red dashed lines represent the second BZ. **j,** Experimental geometry. The mirror plane is parallel to the (11) direction of the sample surface. **k,** Momentum distribution curves (MDCs) at the Fermi level for the 1, 2 and 3UC thin films along cut 1 (shown in panel **g**). Black solid curve is the same MDC extracted from ref. 35. Vertical dashed lines are guides to the eye. **l,** Schematic Fermi surface maps comparing this work and ref. 35.

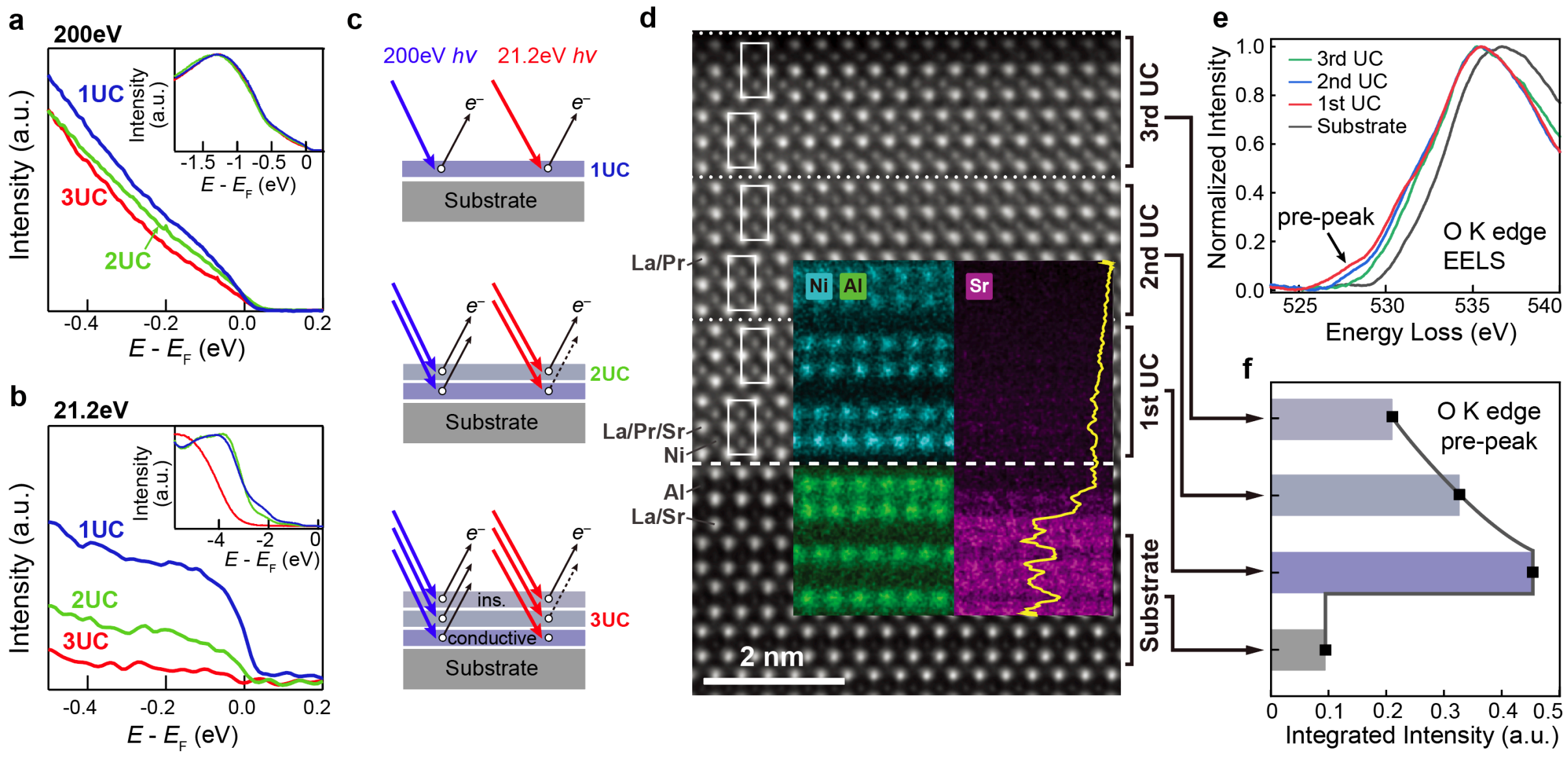

**Figure 2 | Conductivity gradient across the film thickness. a, b,** Momentum-integrated energy distribution curves (EDCs) near the Fermi level of 1, 2 and 3 UC samples using 200 eV synchrotron photons and 21.2 eV Helium-I photons, respectively. Insets show larger energy windows. **c,** Schematics of the possible different photoemission processes across the film thickness under 200 eV and 21.2 eV photons. **d,** High-angle annular dark-field (HAADF) image overlayed with atomically resolved energy-dispersive X-ray spectroscopy (EDS, for Ni, Al and Sr, respectively) images of a 3UC superconducting sample. The yellow curve is the integration of intensity in horizontal pixels for Sr EDS. White dashed line represents the interface. White dotted lines separate different unit cells. White boxes are guides to the eye for the stacking structure of the 327 phase. **e,** Electron energy loss spectroscopy (EELS) of the oxygen K edge, obtained by integrating signals within areas in the substrate and each different UC across a 3UC superconducting sample. The black squared brackets in panel **d** approximately represent the integration windows. Black arrow points to the pre-peak feature. Note that additional oxygen loss may be introduced during STEM sample preparation. **f,** The pre-peak intensity of the curves in panel **e**, integrated from 525 eV to 530 eV. The dark grey solid curve is a schematic profile representing the interfacial confinement.

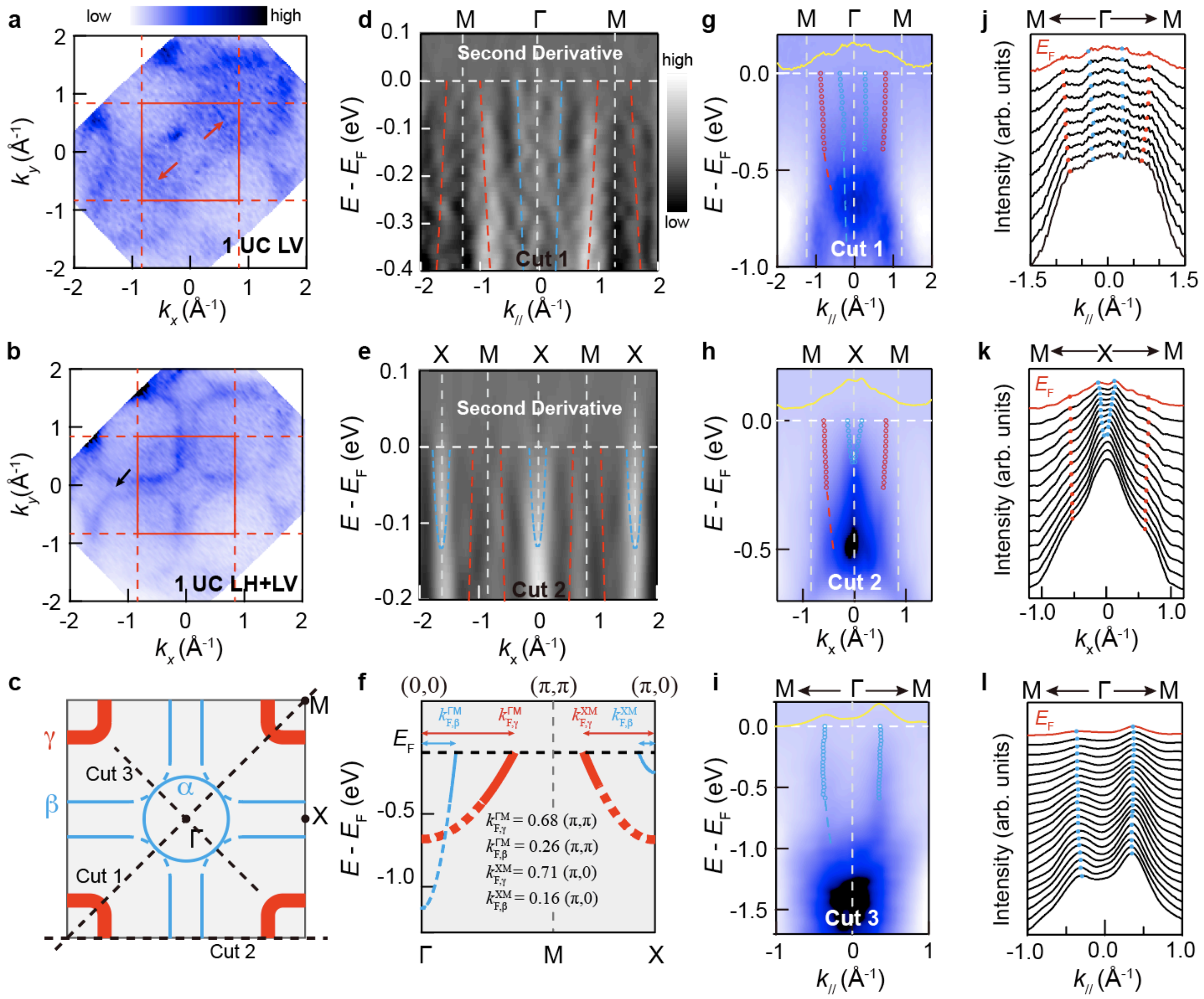

**Figure 3 | Band dispersions. a,** Fermi surface maps of 1UC sample measured by 200 eV LV polarized photons. The red arrows indicate the position of γ pocket dominated by $d_{z^2}$ orbital, which is more clearly resolved in dispersions. **b,** Fermi surface map of the 1UC sample summing intensity obtained from LH and LV photons. The black arrow indicates the small α pocket dominated by $d_{x^2-y^2}$ orbital, which is clearer in the second BZ. Note that in the deflection mode of the hemispheric photoelectron analyzer, polarization analysis deviate from the ideal case at higher angles (e.g. out of the first Brillouin zone). **c,** Schematic of experimentally resolved Fermi surface map with α, β, and γ pockets from different orbital characterizations. The light blue lines represent α and β pockets dominated by Ni 3$d_{x^2-y^2}$ orbital and the bold red lines represent the γ pocket dominated by Ni 3$d_{z^2}$ orbital. **d, e,** Second derivative spectra along cut 1 and 2 in panel **c** with LH photons overlapped with dashed guide lines, respectively. **f,** Schematic of experimentally resolved band dispersion with different orbitals characterized. **g, h,** ARPES spectra along cut 1 and 2 in panel **c** with LH photons, respectively. **i,** ARPES Spectrum along cut 3 with LV photons

extracted from the deflection-mode Fermi surface map as shown in panel **a**. Yellow curves are MDCs at the Fermi level. **j, k, l,** MDC stacks of spectra from panel **g**, **h** and **i** respectively. The color bar shows the ARPES spectra intensity. Blue and red dots mark the position of fitted Lorentzian peaks in MDCs, corresponding to β and γ bands, respectively. Dashed lines are extended guides to the eye.

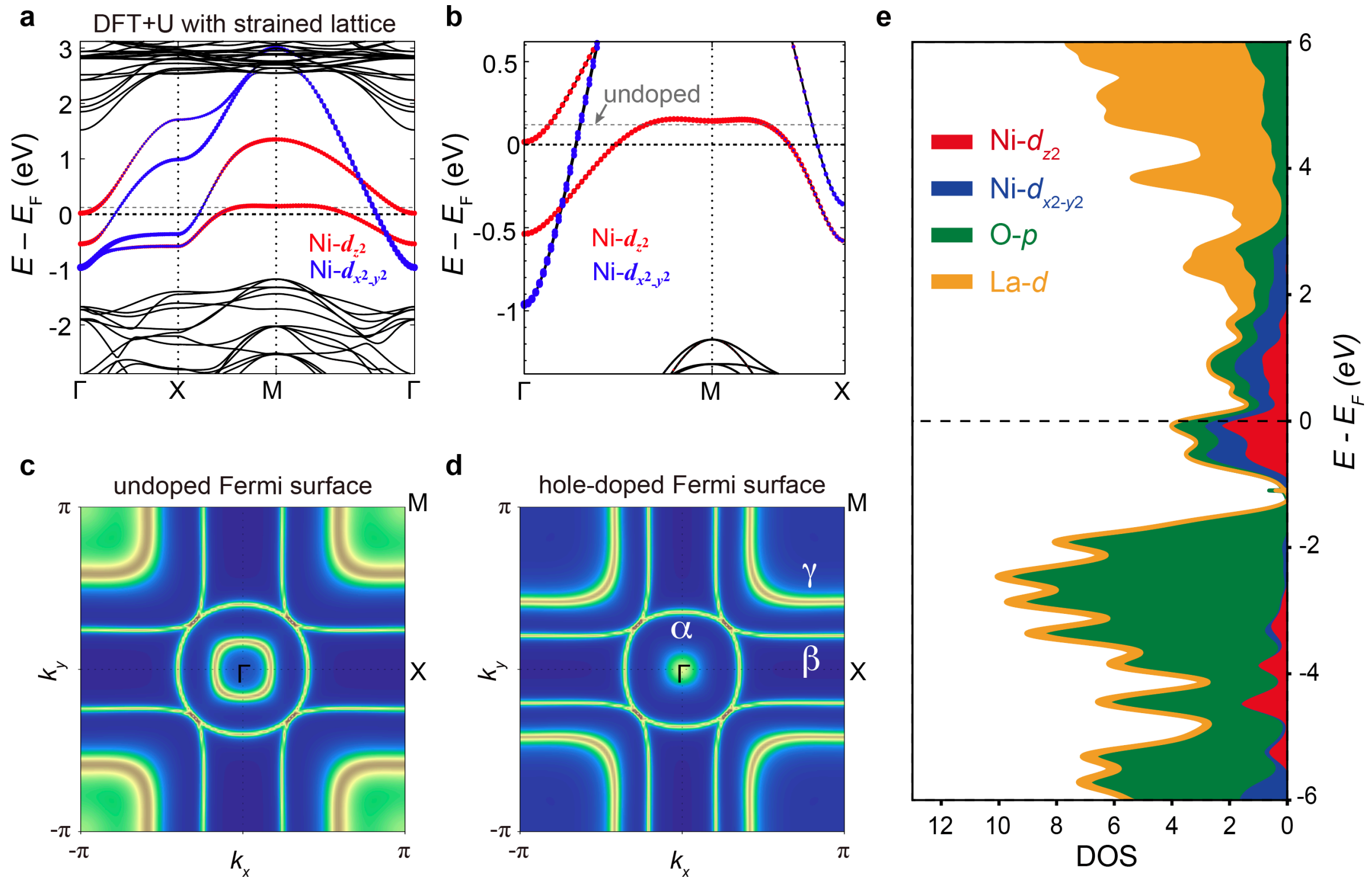

**Figure 4 | DFT+U calculations. a,** Band structures of $La_3Ni_2O_7$ with larger energy window, calculated based on thin film lattice parameters[42]. The red (blue) dots demonstrate the projected weights for the orbital $d_{z^2}$ ($d_{x^2-y^2}$) of Ni atoms. The horizontal grey dashed line indicates the undoped case, while the horizontal black dashed line denotes the hole-doped case corresponding to experiments. **b,** The same data with magnified energy window and high symmetry path similar to Fig. 3f. **c, d,** The Fermi surfaces of undoped and hole-doped cases, respectively. **e,** Orbital-dependent partial density of states.

# Supplementary Materials for

**Angle-resolved photoemission spectroscopy of superconducting $(La,Pr)_3Ni_2O_7/SrLaAlO_4$ heterostructures**

Peng Li[1,2†], Guangdi Zhou[1,2†], Wei Lv[2†], Yueying Li[2†], Changming Yue[1,2,3*], Haoliang Huang[1,2], Lizhi Xu[2], Jianchang Shen[4,5], Yu Miao[4,5], Wenhua Song[2], Zihao Nie[2], Yaqi Chen[2], Heng Wang[2], Weiqiang Chen[1,2], Yaobo Huang[6], Zhen-Hua Chen[6], Tian Qian[7], Junhao Lin[1,2], Junfeng He[4,5*], Yu-Jie Sun[1,2*], Zhuoyu Chen[1,2*], Qi-Kun Xue[1,2,8]

[†]These authors contributed equally.

[*]E-mail: chenzhuoyu@sustech.edu.cn, sunyj@sustech.edu.cn, jfhe@ustc.edu.cn, yuecm@sustech.edu.cn

**a**

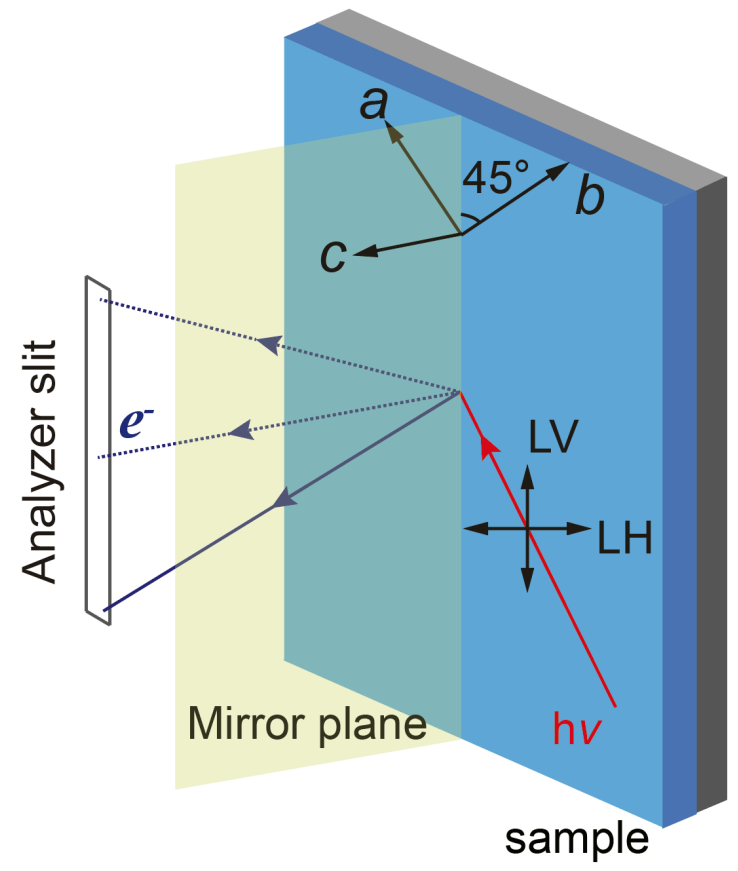

**b**

| | $d_{x^2-y^2}$ | $d_{z^2}$ | $p_x$ | $p_y$ | $p_z$ |
|---|---|---|---|---|---|
| *LV* | × | ○ | ▽ | ▽ | ○ |
| $LH_{perp}$ | ○ | × | ▽ | ▽ | × |
| $LH_{para}$ | × | ○ | ▽ | ▽ | ○ |

○ Allowed ▽ Partially allowed × Forbidden

**Figure S1 Matrix element effect. a**, The geometry of the ARPES measurements with polarized photons at the beamline. The mirror plane is parallel to the (11) direction of the sample surface. **b**, The possibility to detect various Ni-*d* and O-*p* orbitals at the mirror plane with LV and LH photons, based on matrix element analysis. The LV polarization direction is always parallel to the mirror plane. For LH polarized photons, there are two components, one perpendicular (denoted “perp”) to the mirror plane and the other one parallel (denoted “para”) to the mirror plane. The ARPES results from LH photons is the summation of both components, thus allowing more orbitals to be observed.

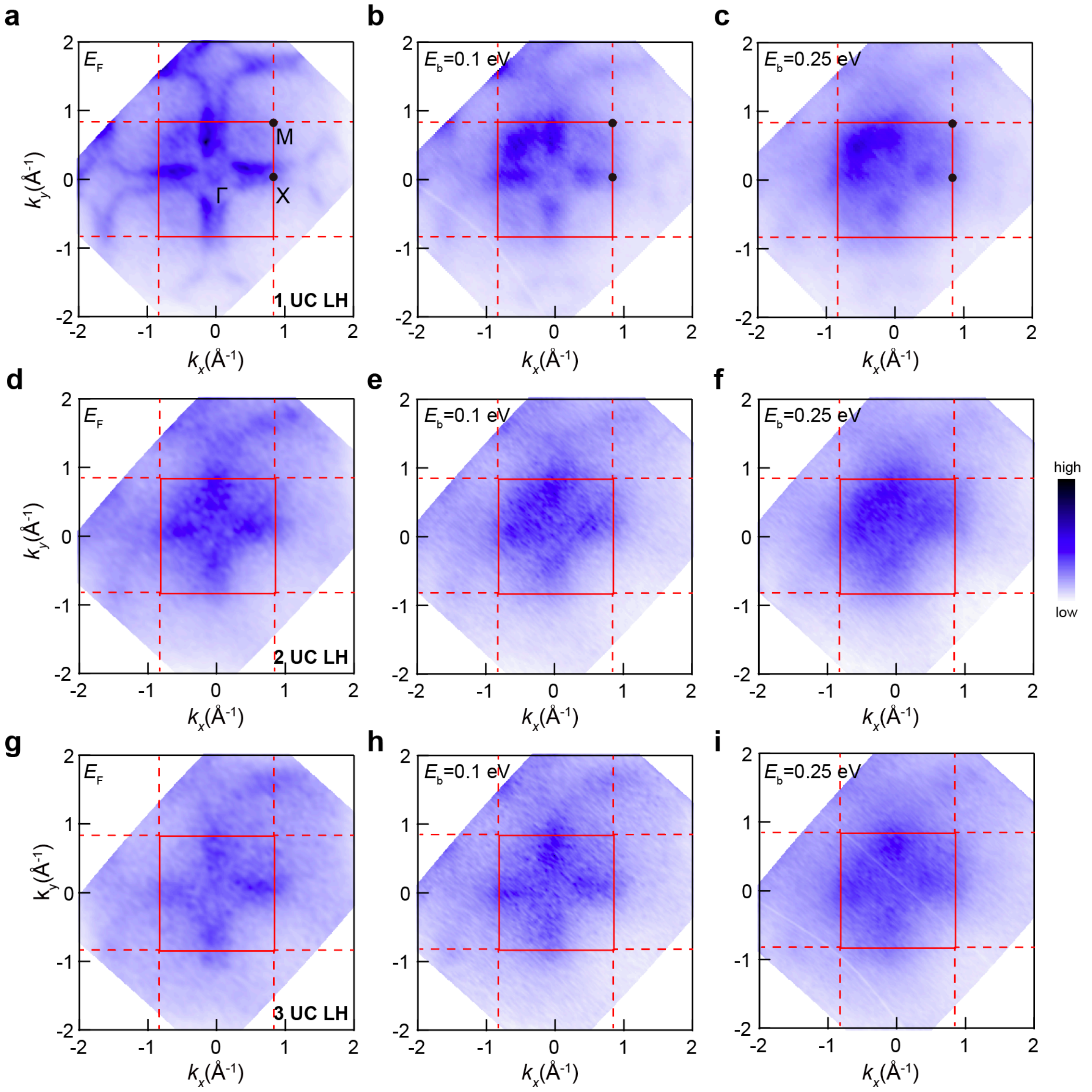

**Figure S2 LH constant energy contours.** Intensity integrated within ±50 meV energy window at specific binding energies with LH polarized photons of 1UC (**a-c**), 2UC (**d-f**) and 3UC (**g-i**) heterostructures. The consistent constant energy contours across 1UC, 2UC, and 3UC further confirm comparable carrier doping level in all three films.

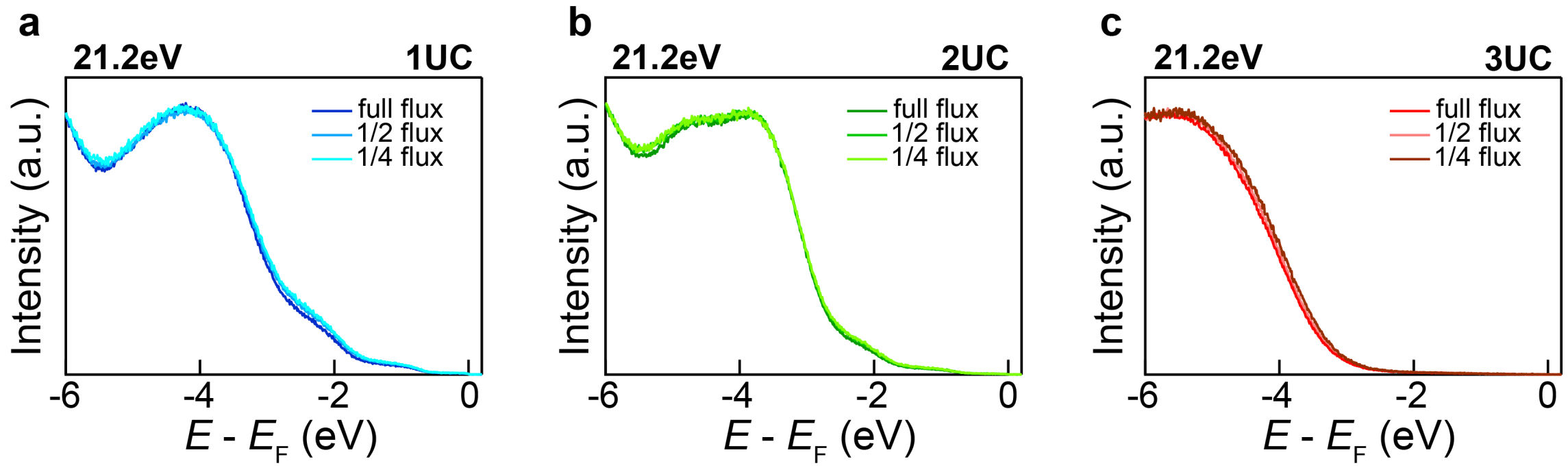

**Figure S3 Flux dependence.** Flux-dependent momentum-integrated energy distribution curves (EDCs) of 1UC (**a**), 2UC (**b**), and 3UC (**c**) samples using 21.2 eV photons. The 1/4, 1/2 and full flux indicate different amounts of injected photons per unit time. If the samples are completely insulating with charging accumulation, the variance of photon flux would lead to different photoelectron intensities and peak positions, which is not the case shown here, implying the existence of conductive channels.

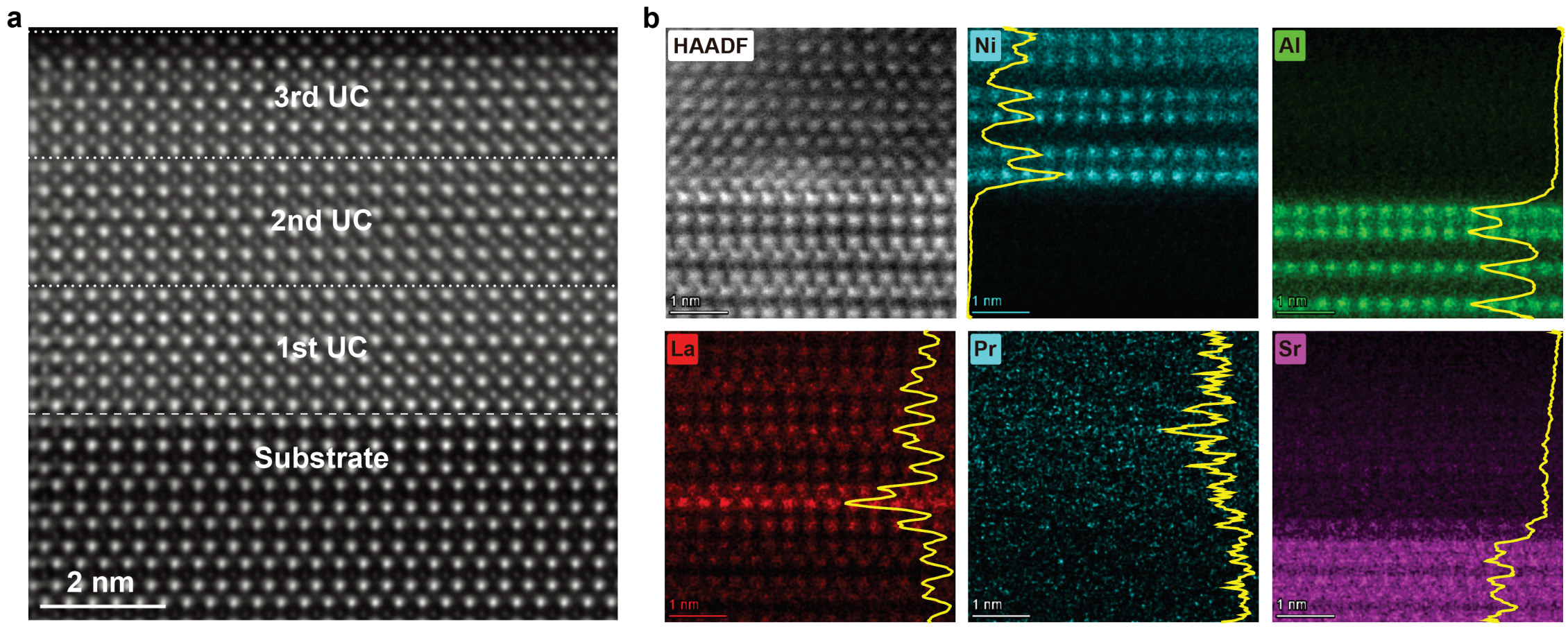

**Figure S4 STEM and EDS.** Same STEM HAADF (**a**) and atomically resolved EDS (**b**) images shown in Fig. 2 in the main text, with larger field of view and added La, Pr data.

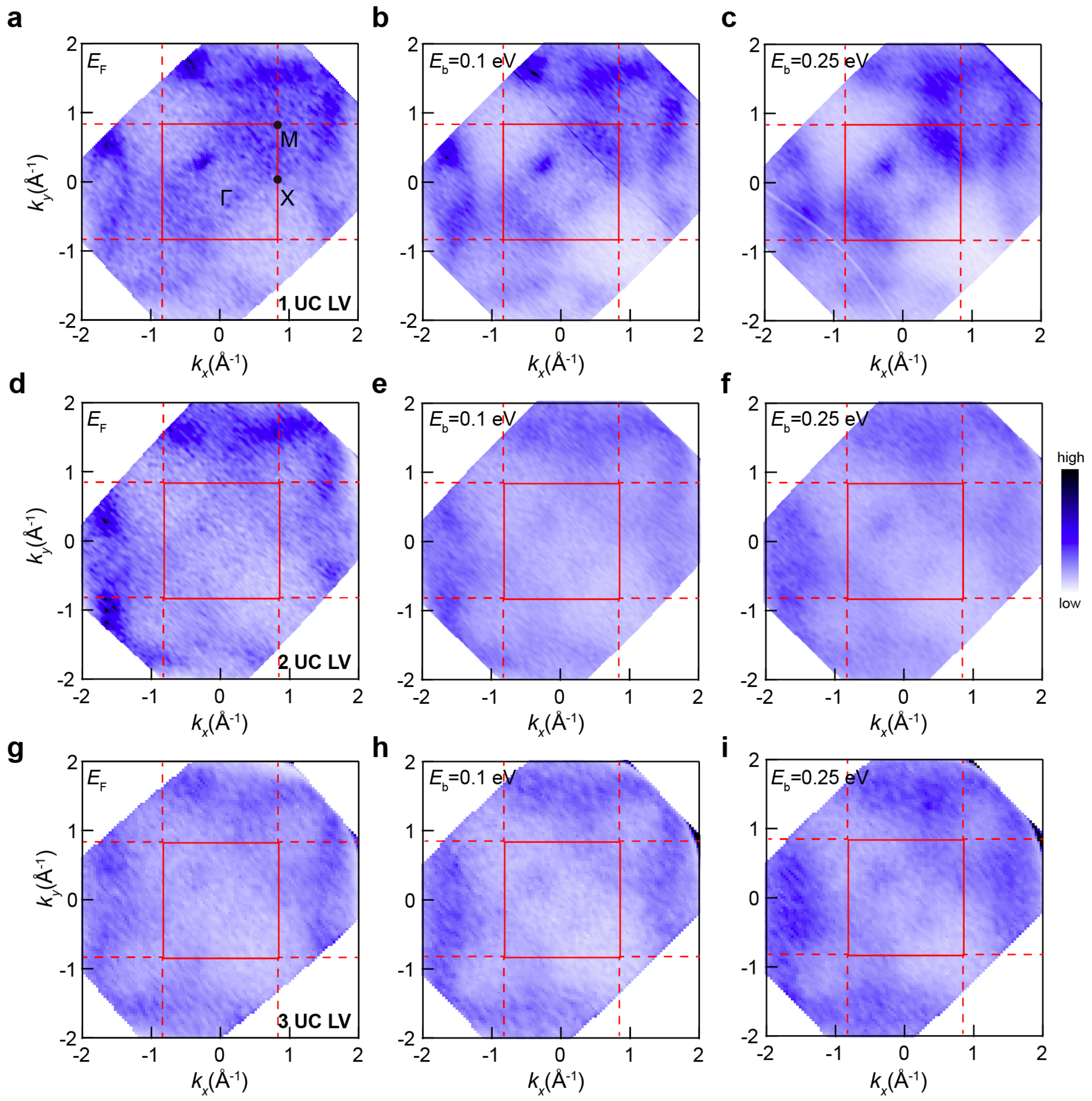

**Figure S5 LV constant energy contours.** Intensity integrated within ±50 meV energy window at specific binding energies with LV polarized photons of 1UC (**a-c**), 2UC (**d-f**) and 3UC (**g-i**) heterostructures. The consistent constant energy contours across 1UC, 2UC, and 3UC further confirm comparable carrier doping level in all three films.

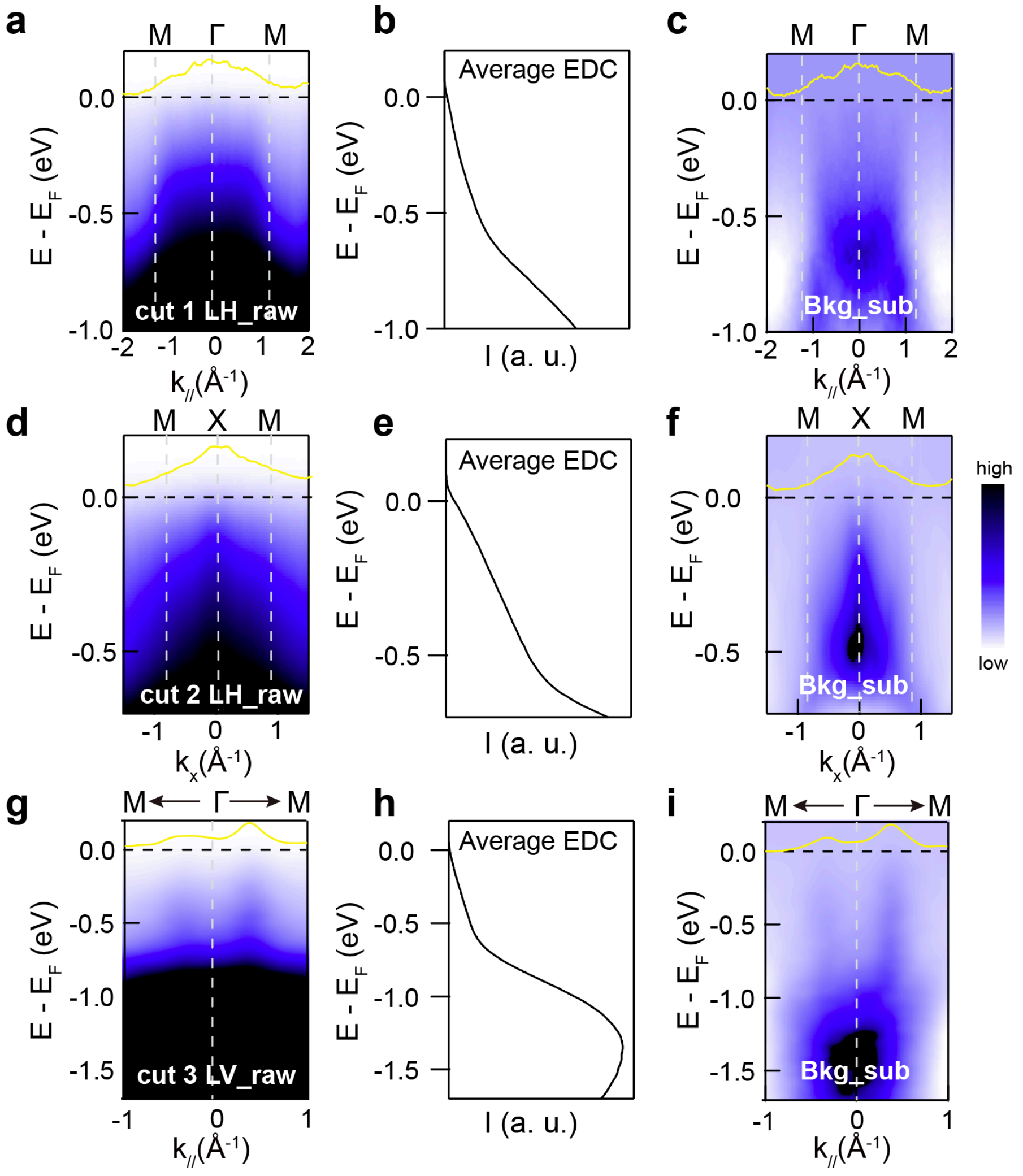

**Figure S6 Background substraction. a, d, g,** Raw data of cuts 1, 2, and 3 in Fig. 3, respectively, with yellow curves representing the raw MDCs at the Fermi level. **b, e, h,** The corresponding average EDCs as the background. **c, f, i,** Spectra after substracting the background to enhance features, same data as the Figs. 3g, 3h and 3i.

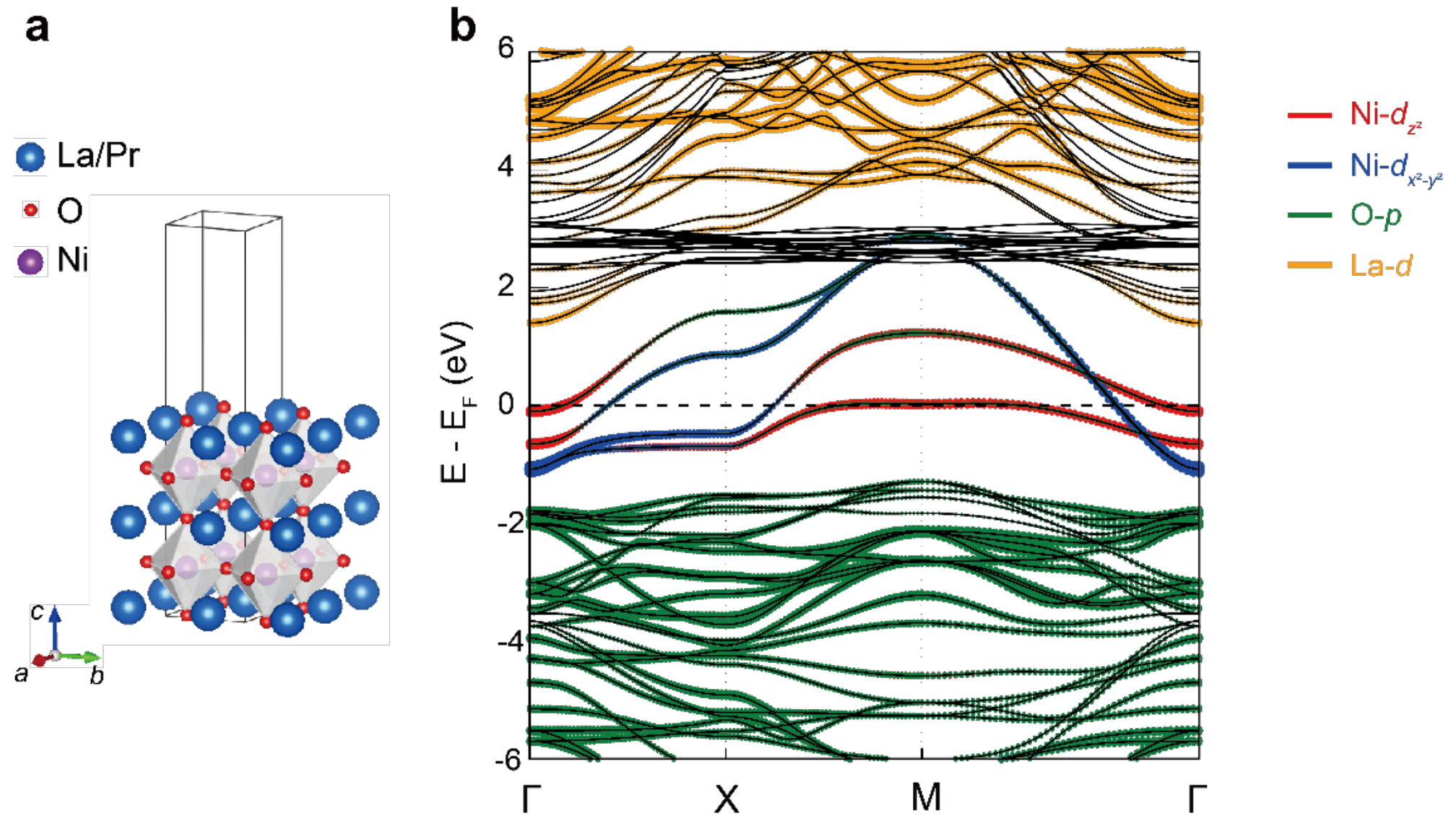

**Figure S7 DFT+U results with orbital characters. a,** The crystal structure of one-bilayer-thick $La_3Ni_2O_7$ thin film constructed with experimentally determined lattice constants [42]. The actual length of vacuum in the calculation is more than 30 Angstrom (much bigger than drawn). **b,** The DFT band structure characterized by different orbitals.